\documentstyle[12pt,axodraw]{article}

\begin{document}

\begin{titlepage}

\hspace{9.5cm}{IFT-P.002/2000}

\vspace{.5cm}

\begin{center}

\LARGE

{\sc Four-Point Amplitude from Open Superstring Field Theory}

\vspace{.5cm}
\large

Nathan Berkovits\footnote{e-mail:  nberkovi@ift.unesp.br} \\ and \\
Carlos Tello Echevarria \footnote{e-mail: carlos@ift.unesp.br} 

\vspace{.5cm}

{\em Instituto de F\'\i sica Te\'orica, Universidade Estadual Paulista} \\
{\em Rua Pamplona 145, 01405-900, S\~ao Paulo, SP, Brasil}

\vspace{.5cm}

December 1999

\end{center}

\vspace{1cm}

\begin{abstract}
An open superstring field theory action has been proposed
which does not suffer from contact term divergences.
In this paper, we compute the on-shell four-point
tree amplitude from this action using the Giddings map.
After including contributions from the quartic term in the action,
the resulting amplitude agrees with the first-quantized prescription.

\end{abstract}

\end{titlepage}

\newpage

\section{Introduction}

To study non-perturbative features of superstring theory, a
field theory action for the superstring could be useful.
As shown by Wendt in 1989 \cite{wendt}, 
Witten's cubic action \cite{witten2} for open superstring field theory
has divergent
contact term problems due to colliding interaction-point operators.
If the Neveu-Schwarz string field
$A$ carries picture $-1$, 
the cubic action is
$\int (A Q A + Z A^3)$ where $Z=\{Q,\xi\}$ is 
the picture-raising operator which is inserted at the interaction-point 
\cite{witten2}.
Since $Z$ has a singularity with itself, the four-point amplitude computed
using the above action has contact term divergences which break
gauge invariance and make the action inconsistent. Similar contact term
problems exist with other choices of picture \cite{preit} and
with other open superstring field theory actions such
as the light-cone and covariantized light-cone actions.\cite{GF}

In 1995, an action for open superstring field theory was proposed which
does not suffer from contact term divergences \cite{field}.
This action was 
constructed by embedding the N=1 superstring into the N=2 superstring
and resembles a
Wess-Zumino-Witten action.
Using modified Green-Schwarz variables, this open
superstring field theory action can be written in a manifestly
SO(3,1) super-Poincar\'e invariant manner \cite{field}. 
This spacetime-supersymmetric
version of the action is easily generalized to any compactification
to four dimensions which preserves at least N=1 $d=4$ supersymmetry.
It is also possible to write this field theory action using 
Ramond-Neveu-Schwarz variables and, at least in the Neveu-Schwarz
sector, this can be done in a manifestly SO(9,1) Lorentz invariant manner.

In this paper, we shall use this action to explicitly compute the
four-point open superstring
tree amplitude. When the external states are on-shell, this
amplitude will be shown to agree with the first-quantized result. There are
no contact term divergences, and the quartic vertex plays a crucial
role in cancelling a {\it finite} contact term contribution coming from
two cubic vertices. 

In section 2, we shall review the 
Wess-Zumino-Witten-like action for open superstring field theory.
In section 3, we shall discuss gauge fixing and the Giddings map
for computing four-point amplitudes using open string field theory.
In section 4, we shall explicitly
compute the four-point on-shell tree amplitude
and will prove equivalence with the first-quantized result.

\section{Review of Superstring Field Theory Action}

For any critical ($\hat c=2$) N=2 superconformal representation with
N=2 generators [$T$,$G^+$,$G^-$,$J$], 
one can construct the following open string field theory action:
\begin{equation}
S=
\frac{1}{2}\int [(e^{-\Phi}G^+_0 e^{\Phi})(e^{-\Phi}\tilde G^+_0 e^{\Phi}) 
-\int_0^1 dt(e^{-t\Phi}\partial_t e^{t\Phi})\left\{e^{-t\Phi}G^+_0 e^{t\Phi},
e^{-t\Phi}\tilde G^+_0 e^{t\Phi}\right\}]. \label{action}
\end{equation}
which resembles the Wess-Zumino-Witten action 
$\int [(g^{-1}dg)^2+
\int dt (g^{-1}dg)^3]$.
The open string fields $\Phi$ are glued together using
Witten's midpoint interaction \cite{witten1}
and are restricted to be U(1)-neutral with
respect to $J$. The open string fields carry Chan-Paton factors which
will be suppressed throughout this paper.

The fermionic operators $G_0^+$ and $\tilde G_0^+$ are constructed 
from the N=2 superconformal generators in the following manner \cite{topo}:
After twisting, the fermionic generator $G^+$ carries conformal weight
$+1$ and the fermionic generator $G^-$ carries conformal weight $+2$.
Furthermore, after bosonizing the U(1) generator $J=\partial H$, 
one can construct SU(2) currents [$J^{++}=e^{iH}, J=\partial H, J^{--}=
e^{-iH}$] of conformal weight $[0,1,2]$ after twisting. Commuting the zero
modes of these
SU(2) currents with the fermionic generators, one obtains two new fermionic
generators, $\tilde G^+=[G^-,J_0^{++}]$ and $\tilde G^-=[G^+,J_0^{--}]$,
of conformal weight $+1$ and $+2$ respectively. 
$G_0^+$ is defined to be the zero mode of $G^+$ and $\tilde G^+_0$
is defined to be the zero mode of $\tilde G^+$. Note that 
$\{G^+_0,\tilde G^+_0\}=0$ and after twisting, the U(1) anomaly
implies that non-vanishing correlation functions on a sphere must
carry $+2$ U(1) charge.

The action of (\ref{action})
can be justified by considering the following three examples
of critical N=2 superconformal representations. The first example is
the open self-dual string where $\Phi$ depends on the string modes
of the variables 
$[x^{\alpha\dot\alpha}, 
\psi^\alpha, \bar\psi^{\alpha}]$ for $\alpha,\dot\alpha$
=1 to 2. For this N=2 superconformal representation, 
$G^+ =\psi_{\alpha}\partial x^{\alpha +}$,
$\tilde G^+ =\psi_{\alpha}\partial x^{\alpha -}$, and $J=\psi^\alpha
\bar\psi_{\alpha}$. Since there are no massive physical fields,
the dependence of $\Phi$ on the non-zero modes
of
$[x^{\alpha\dot\alpha}, \psi^\alpha, \bar\psi^{\alpha}]$ gives rise
to auxiliary and gauge fields. The massless physical field comes
from dependence on the zero mode of $x^{\alpha\dot
\alpha}$, $\Phi (x_0^{\alpha\dot\alpha})$, which when plugged into
(\ref{action}) reproduces
the Donaldson-Nair-Schiff action \cite{Donaldson} for
D=4 self-dual Yang-Mills.

The second example is the N=2 embedding of the superstring using 
the standard RNS worldsheet variables $[x^\mu,\psi^\mu,b,c,\xi,\eta,\phi]$
where $\mu=0$ to 9 and
the super-reparameterization ghosts have been fermionized as
$\beta= e^{-\phi}\partial\xi$ and $\gamma= e^\phi\eta$ \cite{topo}.
In this example,
the Neveu-Schwarz contribution to the field theory action is obtained
by defining
$\Phi =\xi_0 A$ where $A$ is the standard Neveu-Schwarz string field in the
$-1$ picture. (The Ramond contribution to the superstring field theory
action is not yet known.) In the N=2 embedding of the superstring, $G_0^+$
is the BRST charge $Q$ and $\tilde G_0^+$ is the zero mode of $\eta$. 
So the linearized equation of motion coming from the action of (\ref{action}) 
is
$$0 = G_0^+ \tilde G_0^+ \Phi = Q \eta_0 (\xi_0 A) = Q A$$
as desired. Furthermore, since 
$J=bc +\xi\eta$, the U(1)-neutrality condition implies that
$\Phi$ has zero RNS ghost-number.

The third example is the `hybrid' description of the superstring using
the four-dimensional superspace variables $[x^m,\theta^\alpha,
\bar\theta^{\dot\alpha}]$ where $m$=0 to 3 and $\alpha,\dot\alpha=1$ to 2,
combined with a chiral boson $\rho$ and the variables of a $c=9$
N=2 superconformal field theory which describes a six-dimensional
compactification manifold \cite{four}.
In this example, the compactification-independent part of the
field theory action is obtained by defining
$\Phi$ to be a string
field depending only on the four-dimensional superspace variables.
(The compactification-dependent contribution to the action
was worked out in \cite{field} and involves two other string fields in
addition to $\Phi$.)
The massless and first massive level of this action have been 
explicitly computed in \cite{field} and \cite{leite} 
and describe a super-Yang-Mills
and massive spin-2 multiplet in N=1 $d=4$ superspace.

\section{Gauge Fixing and the Giddings Map}

Similar to the Wess-Zumino-Witten action, the action of (\ref{action})
is invariant
under the gauge transformation
$$\delta e^\Phi = (G_0^+\Omega) e^\Phi + e^\Phi(\tilde G_0^+\tilde\Omega)$$
where $\Omega$ and $\tilde\Omega$ are arbitrary string fields of
$-1$ U(1)-charge. At linearized level, this transformation reduces
to $\delta \Phi = G_0^+\Omega + \tilde G_0^+\tilde\Omega$
which allows the gauge-fixing conditions 
\begin {equation}
G_0^-\Phi=\tilde G_0^-\Phi=0 \label{gaugefixing}
\end {equation}
where $G_0^-$ and $\tilde G_0^-$ are the zero modes of $G^-$ and
$\tilde G^- =[G^+,J_0^{--}]$. 
The gauge-fixing conditions of (\ref{gaugefixing}) 
can be obtained, for example, by
choosing 
$$\Omega = -(T_0)^{-1} G^-_0 \Phi, \quad \tilde\Omega=
-(T_0)^{-2} \tilde G_0^-  G_0^- G_0^+ \Phi,$$ 
so that
$G^-_0\delta\Phi= -G^-_0\Phi$ and 
$\tilde G^-_0\delta\Phi= -\tilde G^-_0\Phi$
where we have used that
$\{G_0^-,G_0^+\}
=\{\tilde G_0^-,\tilde G_0^+\}=T_0$.

In this gauge, the linearized propagator is 
${\cal P}=(T_0)^{-2} G^-_0\tilde G^-_0$
since 
$\tilde G^+_0 G^+_0 \Phi$
$ = 0$ is the linearized equation of motion
and $ {\cal P}\tilde G^+_0 G^+_0 \Phi = \Phi$. 
Although this propagator looks complicated, we shall show in 
section 4 that
it can be simplified to 
${\cal P}=(T_0)^{-1} J^{--}_0$ when computing on-shell tree amplitudes.

Although the action of (1) 
contains vertices with arbitrary numbers of string fields, only
the cubic and quartic vertices will be necessary for computing four-point
tree amplitudes. Expanding the action of (1), one obtains 
\begin{equation}
S=\int (\frac{1}{2}G^+_0 \Phi \tilde G^+_0 \Phi 
- \frac{1}{6}\Phi \{G^+_0\Phi,\tilde G^+_0\Phi\}
-\frac{1}{24}[\Phi ,G^+_0\Phi][\Phi , \tilde G^+_0\Phi]+\cdots ).
\label{vertices}
\end{equation}
Using Witten's gluing prescription \cite{witten1}
for string fields, the cubic and quartic vertex from  (\ref{vertices})
are described by the diagrams

\begin{center}
\begin{picture}(360,100)(0,0)
\Line(20,20)(80,20)  \Text(20,60)[]{$(3)$}
\Line(20,20)(70,70)
\Line(70,70)(130,70)
\Line(130,70)(110,50) \Text(150,60)[]{$(2)$}
\Line(100,40)(80,20) \Text(120,30)[]{$(1)$}
\Line(100,40)(70,45)
\DashLine(70,45)(110,50){3} \Text(40,30)[]{$ $}
\Line(100,40)(100,60)
\Line(100,60)(70,65)
\Line(70,65)(110,70)
\Line(110,70)(110,50)
\Line(70,45)(70,65)

\Line(220,20)(280,20)  %%inferiorline
\Line(270,70)(330,70)    %%superior-line
\Line(330,70)(310,50)  \Text(350,60)[]{$(2)$} %%sup/
  \DashLine(270,70)(250,50){3}  \Text(230,70)[]{$(3)$} %sup/
\Line(300,40)(280,20)  \Text(320,30)[]{$(1)$} %inf/
   \Line(240,40)(220,20) \Text(200,40)[]{$(4)$}%inf/
\Line(300,40)(270,45)  %%inf/-center   ***
   \Line(240,40)(270,45) \Text(240,30)[]{$ $}
\DashLine(270,45)(310,50){3}  %% center-sup/
    \DashLine(270,45)(250,50){3}
\Line(300,40)(300,60) % vert inf/
    \Line(240,40)(240,60)
\Line(300,60)(271,65) % inf/-center up  ***
    \Line(240,60)(269,65)   %***
\Line(271,65)(310,70)  % sup/-center up ***
    \Line(269,65)(250,70)       %%%***
\Line(310,70)(310,50)  % vert sup
   \Line(250,70)(250,50)
\Line(270,45)(269,65) \Line(270,45)(271,65)  % vert center ****
   \Text(40,0)[]{}\Text(260,0)[]{}
\end{picture} \\
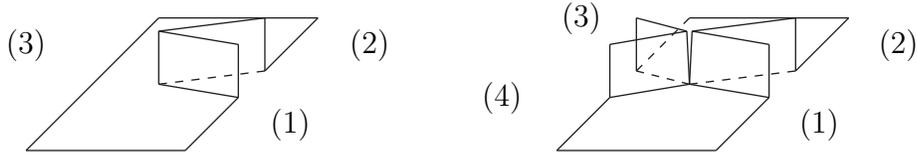
{\sl Figure 1}
\end{center}

For four-point tree amplitudes, contributions can come from two cubic
vertices or from one
quartic vertex.
The Witten diagram for two cubic vertices connected by a propagator
of length $\tau$ is given by
\begin{center}
\begin{picture}(140,100)(0,0)
%\Line(180,70)(180,90)
%\Line(200,70)(180,70)
%\Text(195,85)[]{w}
\Line(0,0)(0,50)
\Line(0,50)(140,50)
\Line(140,50)(140,0)
\Line(140,0)(90,0)
\Line(90,0)(90,25)
\Line(90,25)(50,25)
\Line(50,25)(50,0)
\Line(50,0)(0,0)
\DashLine(3,3)(3,53){2}
\Line(3,53)(143,53)
\Line(143,53)(143,3)
\DashLine(143,3)(90,0){2}
\DashLine(50,0)(3,3){2}
\Text(-10,25)[]{$(1)$}
\Text(10,15)[]{$(2)$}
\Text(150,25)[]{$(4)$}
\Text(130,15)[]{$(3)$}

%\Text(95,5)[]{F}
%\Text(95,25)[]{E}
%\Text(45,5)[]{C}
%\Text(45,25)[]{D}

\Line(50,-5)(50,-15)\Line(90,-5)(90,-15)
\Text(55,-12)[]{$\leftarrow$}
\Text(85,-12)[]{$\rightarrow$}
\Text(70,-12)[]{$\tau$}

\Line(160,0)(170,0)\Line(160,50)(170,50)
\Text(165,5)[]{$\downarrow$}
\Text(165,45)[]{$\uparrow$}
\Text(165,25)[]{$\pi$}

\Line(-30,0)(-20,0)\Line(-30,50)(-20,50)
\Text(-25,5)[]{$\downarrow$}
\Text(-25,45)[]{$\uparrow$}

\Text(-25,25)[]{$\pi$}

\Line(60,0)(70,0)
\Text(65,5)[]{$\downarrow$}
\Text(65,20)[]{$\uparrow$}
\Text(65,12)[]{$\pi/2$}

\Line(70,25)(68,50)\DashLine(70,25)(72,53){3}
\Text(63,36)[]{c} \Text(150,-20)[] {}
\end{picture} \\
\vskip 1cm
{\sl Figure 2}
\end{center}
where the propagator is integrated along the contour $c$.
The Witten diagram for a quartic vertex 
is given by Figure 2 in the limit that $\tau\to 0$.

As was shown by Giddings \cite{giddings}, it is convenient to perform
a Schwarz-Christoffel transformation from the diagram of figure 2
to the upper half plane such that the four external strings are mapped
to the points $\pm\alpha(\tau)$ and $\pm (\alpha(\tau))^{-1}$. 
If strings $[(1),(2),(3),(4)]$ 
are mapped to the points $[-\alpha, \alpha,$
$ \alpha^{-1}, -\alpha^{-1}]$
in this order, then  
$0<\alpha(\tau)\leq \delta$ where $\delta\equiv
\sqrt{2}-1$, $\alpha(\tau=0)=\delta$ and $\alpha(\tau=\infty)=0$. 
On the other hand, if strings $[(1),(2),(3),(4)]$ 
are mapped to the points $[-\alpha^{-1}, -\alpha, \alpha, \alpha^{-1}]$
in this order, then  
$\delta\leq\alpha(\tau)<1$
where 
$\alpha(\tau=0)=\delta$ and $\alpha(\tau=\infty)=1$. 

\section{Computation of Four-Point Amplitude }

The tree-level scattering amplitude for
four external string fields labeled by the letters $[A,B,C,D]$
gets contributions either from the diagram
of Figure 2 or from the second
diagram of Figure 1. There are 24 different ways to match
the string fields $[A,B,C,D]$ with the external legs $[(1),(2),(3),(4)]$
and we shall restrict our attention to those 4 combinations where 
$[A,B,C,D]$ are cyclically ordered. For contributions coming from the
diagram of Figure 2, these 4 combinations split into two `s-channel'
contributions, ${\cal A}_s$, 
where string $A$ is associated with leg (2) or leg (4) and
two `t-channel' contributions,
${\cal A}_t$, where string $A$ is associated with leg (1) or
leg (3). The four cyclically related combinations
coming from the quartic
vertex of Figure 1 will be called ${\cal A}_q$.

In this section, it will be shown that the sum of these three
contributions,
${\cal A}={\cal A}_s +{\cal A}_t + {\cal A}_q$,
reproduces the first-quantized result for the on-shell
scattering amplitude which can be written as \cite{topo}
\begin{equation}
{\cal A} =  \int_0^1 d\alpha ~\langle ~~(\int d^2 z \mu_\alpha
(z,\bar z) G^-(z)) 
\label{first}
\end{equation}
$$\Phi_A(-\alpha^{-1})~ G^+_0 \Phi_B(-\alpha)
~G^+_0 \Phi_C(\alpha) ~\tilde G^+_0 \Phi_D(\alpha^{-1}) \rangle. $$
where $\mu_{\alpha}(z,\bar z)$ is the appropriate Beltrami differential for an
$\alpha$-dependent parameterization of the modulus and $\langle~\rangle$
signifies the two-dimensional correlation function in the upper half-plane. 
To relate (\ref{first}) to the standard expression for the scattering
amplitude
of four Neveu-Schwarz strings, recall from section 2 that 
$\Phi =\xi_0 V^{(-1)}$ where $V^{(-1)}$
is the Neveu-Schwarz vertex operator in the $-1$ picture. 
Furthermore, $G^- =b$,
$\tilde G^+_0 \Phi = \eta_0 \xi_0 V^{(-1)} = V^{(-1)}$ and 
$G^+_0 \Phi = Q \xi_0 V^{(-1)} = V^{(0)}$ 
where $V^{(0)}$ is the 
Neveu-Schwarz vertex operator in the zero picture. 
So (\ref{first})
implies that 
$$
{\cal A} =  \int_0^1 d\alpha \langle (\int d^2 z \mu_\alpha(z,\bar z) b(z)) 
\xi_0 V^{(-1)}_A(-\alpha^{-1}) ~  V_B^{(0)}(-\alpha)~
V^{(0)}_C(\alpha) ~V^{(-1)}_D(\alpha^{-1})\rangle,$$
which agrees 
with the first-quantized
prescription of \cite{FMS} 
in the large Hilbert 
space, i.e. in the
Hilbert space including the $\xi$ zero mode.

\subsection{Computation of s-channel contribution}

We shall first compute the contribution ${\cal A}_s$ from s-channel diagrams.
Using the cubic vertices and propagator ${\cal P}$
from section 3 and including the various combinatorical factors,
$${\cal A}_s = 2 (3)^2 \frac{1}{2}(-\frac{1}{6})^2 
\langle (~ G^+_0\Phi_A(4)~\tilde G^+_0\Phi_B(1) + 
\tilde G^+_0\Phi_A(4)~ G^+_0\Phi_B(1) ) $$
$${\cal P} ~
( G^+_0\Phi_C(2)~\tilde G^+_0\Phi_D(3) + 
\tilde G^+_0\Phi_C(2)~ G^+_0\Phi_D(3) )~ \rangle_W $$
where $\langle~\rangle_W$ signifies the two-dimensional correlation
function in the Witten diagram of Figure 2.
Note that one can always choose
the $G^+_0$ and $\tilde G_0^+$ operators of the cubic vertices
to act on external legs since
$$\langle ~(G^+_0 \Phi_A(1) ~\tilde G^+_0 \Phi_B(2) +
\tilde G^+_0 \Phi_A(1)~ G^+_0 \Phi_B(2) ) ~\Phi_C(3)~\rangle =$$
$$
\langle~  \Phi_A(1)~ (\tilde G^+_0 \Phi_B(2)~ G^+_0 \Phi_C(3) 
+ G^+_0 \Phi_B(2)~ \tilde G^+_0\Phi_C(3))~\rangle$$
by deforming the contour integral of $G^+$ and $\tilde G^+$ off of
$\Phi_A(1)$. 

The propagator ${\cal P} = (T_0)^{-2} G^-_0 \tilde G^-_0$ can
be simplified by writing $\tilde G^-_0 = [G^+_0, J_0^{--}]$ and
deforming the contour integral of $G^+$ off of $J^{--}_0$.
When the external string fields are on-shell, i.e. $G^+_0 \tilde G^+_0
\Phi =0$, the contour integral of $G^+$ 
only contributes by hitting the remaining
$G^-_0$ of the propagator. Since $\{G^+_0,G^-_0\} = T_0$, this means
that the propagator can be simplified to ${\cal P} = (T_0)^{-1} J_0^{--}$
when all external string fields are on-shell.
As usual, it is convenient to rewrite 
$(T_0)^{-1} = \int_0^\infty d\tau e^{-\tau T_0}$ 
so the s-channel contribution is
$${\cal A}_s =  \frac{1}{4}\int_0^\infty d\tau 
\langle ~\int_{c}\frac{dw}{2\pi i} J^{--}(w)
 ( G^+_0\Phi_A(4)~\tilde G^+_0\Phi_B(1) + 
\tilde G^+_0\Phi_A(4)~ G^+_0\Phi_B(1) ) $$
$$
 ( G^+_0\Phi_C(2)~\tilde G^+_0\Phi_D(3) + 
\tilde G^+_0\Phi_C(2)~ G^+_0\Phi_D(3) ) ~\rangle_W $$
where the contour $c$ is that of Figure 2.

After performing the Giddings map of Figure 2 to the upper half plane 
using the ordering $[(1),(2),(3),(4)]\to [-\alpha,\alpha,
\alpha^{-1}, -\alpha^{-1}]$, 
\begin{equation}
{\cal A}_s = - \frac{1}{4} 
\int_0^{\delta} d \alpha (\frac{d\tau}{d\alpha}) 
\langle \int_{\tilde c}\frac{dz}{2\pi i}(\frac{dz}{dw}) J^{--}(z) 
\label{As}
\end{equation}
$$( G^+_0\Phi_A(-\alpha^{-1})~\tilde G^+_0\Phi_B(-\alpha) + 
\tilde G^+_0\Phi_A(-\alpha^{-1})~ G^+_0\Phi_B(-\alpha) ) $$
$$( G^+_0\Phi_C(\alpha)~\tilde G^+_0\Phi_D(\alpha^{-1}) + 
\tilde G^+_0\Phi_C(\alpha)~ G^+_0\Phi_D(\alpha^{-1}) ) \rangle $$
where $\tilde c$ is the Giddings map of the contour $c$ and we have
used that $J^{--}$ has conformal weight $2$ so
$J^{--}(z) = (\frac{dz}{dw})^2 J^{--}(w)$.
The overall minus sign comes from the fact that $\alpha$ decreases
as $\tau$ increases.

\subsection{Computation of t-channel contribution}

Performing a re-identification
of the external strings with the external legs,
one finds that the t-channel contribution to the scattering amplitude is
$$
{\cal A}_t =  \frac{1}{4}\int_0^\infty d\tau 
\langle\int_{c}\frac{dw}{2\pi i} J^{--}(w) 
$$
$$( G^+_0\Phi_D(4)~\tilde G^+_0\Phi_A(1) + 
\tilde G^+_0\Phi_D(4)~ G^+_0\Phi_A(1) ) $$
$$( G^+_0\Phi_B(2)~\tilde G^+_0\Phi_C(3) + 
\tilde G^+_0\Phi_B(2)~ G^+_0\Phi_C(3) ) \rangle_W $$
\begin{equation}
= \frac{1}{4} 
\int_{\delta}^1 d \alpha (\frac{d\tau}{d\alpha}) 
\langle \int_{\tilde c}\frac{dz}{2\pi i}(\frac{dz}{dw}) J^{--}(z) 
\label{At}
\end{equation}
$$( G^+_0\Phi_D(\alpha^{-1})~\tilde G^+_0\Phi_A(-\alpha^{-1}) + 
\tilde G^+_0\Phi_D(\alpha^{-1})~ G^+_0\Phi_A(-\alpha^{-1}) ) $$
$$( G^+_0\Phi_B(-\alpha)~\tilde G^+_0\Phi_C(\alpha) + 
\tilde G^+_0\Phi_B(-\alpha)~ G^+_0\Phi_C(\alpha) ) \rangle $$
where we have now
used the Giddings map with the ordering
$[(1),(2),(3),(4)]$
$\to [-\alpha^{-1},-\alpha,
\alpha, \alpha^{-1}]$, 

If there were no $G^+_0$ and $\tilde G^+_0$ operators, 
one could sum ${\cal A}_s$ and ${\cal A}_t$ to get an integral
$\int_0^1 d\alpha f(\alpha)$.
This is what happens in the 
open bosonic string four-point amplitude where
$\int_0^1 d\alpha f(\alpha)$ can be related to the Veneziano amplitude
using the fact that 
$$(\frac{d\tau}{d\alpha}) 
\int_{\tilde c}\frac{dz}{2\pi i}(\frac{dz}{dw}) b(z)=
\int d^2 z \mu_\alpha(z,\bar z) b(z)$$ where $\mu_{\alpha}$
is the Beltrami differential corresponding to the $\alpha$ modulus
\cite{giddings}.
However, for the open superstring four-point amplitude,
one first has to perform contour deformations of
the $G^+_0$ and $\tilde G^+_0$ operators in ${\cal A}_t$ until they 
appear in the same
manner as in the expression for ${\cal A}_s$. 
As will now be shown, these contour deformations produce a {\it finite}
contact term which is cancelled by the contribution from the quartic
vertex.

Consider the expression
$$   
Y=\frac{d\tau}{d\alpha}\langle  
\int_{\tilde c}\frac{dz}{2\pi i}(\frac{dz}{dw}) J^{--}(z)
G^+_0\Phi_D(\alpha^{-1})~\tilde G^+_0\Phi_A(-\alpha^{-1})
~\tilde G^+_0\Phi_B(-\alpha)~ G^+_0\Phi_C(\alpha)  \rangle $$
$$   
=-\langle  
(\int d^2 z \mu_{\alpha}(z,\bar z) J^{--}(z))
~\tilde G^+_0\Phi_A(-\alpha^{-1})
\tilde G^+_0\Phi_B(-\alpha) G^+_0\Phi_C(\alpha)  
G^+_0\Phi_D(\alpha^{-1})
~\rangle $$
which is one of the terms in ${\cal A}_t$. Deforming the $G^+$ contour off
of $\Phi_D(\alpha^{-1})$, one gets
$$Y=
-\langle  
(\int d^2 z \mu_{\alpha}(z,\bar z) \tilde G^{-}(z))
~\tilde G^+_0\Phi_A(-\alpha^{-1}) 
~\tilde G^+_0\Phi_B(-\alpha)~ G^+_0\Phi_C(\alpha) 
~\Phi_D(\alpha^{-1})~ \rangle.$$
Now deform the contour of $\tilde G^+$ off of $\Phi_B(-\alpha)$
to get
$$Y=
-\langle 
~(\int d^2 z \mu_{\alpha}(z,\bar z) \tilde G^{-}(z))
~ \tilde G^+_0
\Phi_A(-\alpha^{-1}) 
~\Phi_B(-\alpha)~ G^+_0\Phi_C(\alpha)~  
\tilde G_0^+\Phi_D(\alpha^{-1})~ \rangle$$
$$+\langle 
~(\int d^2 z \mu_{\alpha}(z,\bar z) T(z))
~\tilde G^+_0\Phi_A(-\alpha^{-1}) 
~\Phi_B(-\alpha)~ G^+_0\Phi_C(\alpha)  
~\Phi_D(\alpha^{-1})~ \rangle$$
where $T$ is the stress-tensor. Finally, writing $\tilde G^-(z) =
[G^+_0, J^{--}(z)]$ and deforming the $G^+$ contour off of $J^{--}(z)$,
one gets 
\begin{equation}
Y=
-\langle 
~(\int d^2 z \mu_{\alpha}(z,\bar z) J^{--}(z))
\label{Ydef}
\end{equation}
$$ \tilde G^+_0\Phi_A(-\alpha^{-1}) 
~ G^+_0\Phi_B(-\alpha)~ G^+_0\Phi_C(\alpha) ~
~ \tilde G^+_0\Phi_D(\alpha^{-1})~ \rangle $$
$$+ \frac{\partial}{\partial\alpha} 
\langle \tilde G_0^+\Phi_A(-\alpha^{-1})~\Phi_B(-\alpha) 
~ G^+_0\Phi_C(\alpha)~ \Phi_D(\alpha^{-1})  \rangle $$
where we have used that 
$\int d^2 z \mu_{\alpha}(z,\bar z) T(z)$
has the effect of taking the variation
with respect to the modulus $\alpha$.

Similarly, 
\begin{equation}
\frac{d\tau}{d\alpha}
\langle
\int_{\tilde c}\frac{dz}{2\pi i}(\frac{dz}{dw}) J^{--}(z)
~\tilde G^+_0\Phi_D(\alpha^{-1})~ G^+_0\Phi_A(-\alpha^{-1})
~G^+_0\Phi_B(-\alpha)~ \tilde G^+_0\Phi_C(\alpha)  \rangle
\label{similar}
\end{equation}
$$=
-\langle
~(\int d^2 z \mu_{\alpha}(z,\bar z) J^{--}(z))$$
$$   G^+_0\Phi_A(-\alpha^{-1})
~\tilde G^+_0\Phi_B(-\alpha)~ \tilde G^+_0\Phi_C(\alpha) ~
~ G^+_0\Phi_D(\alpha^{-1}) \rangle $$
$$- \frac{\partial}{\partial\alpha}
\langle G_0^+\Phi_A(-\alpha^{-1})~\Phi_B(-\alpha)
~ \tilde G^+_0\Phi_C(\alpha)~ \Phi_D(\alpha^{-1})  \rangle. $$

Plugging (\ref{Ydef}) and (\ref{similar}) into (\ref{At}), one
finds
\begin{equation}
{\cal A}_t = - \frac{1}{4} 
\int_{\delta}^1 d \alpha 
\langle~(\int d^2 z \mu_{\alpha}(z,\bar z)  J^{--}(z)) 
\label{Attwo}
\end{equation}
$$( G^+_0\Phi_A(-\alpha^{-1})~\tilde G^+_0\Phi_B(-\alpha) + 
\tilde G^+_0\Phi_A(-\alpha^{-1})~ G^+_0\Phi_B(-\alpha) ) $$
$$( G^+_0\Phi_C(\alpha)~\tilde G^+_0\Phi_D(\alpha^{-1}) + 
\tilde G^+_0\Phi_C(\alpha)~ G^+_0\Phi_D(\alpha^{-1}) ) \rangle $$
$$+ 
\frac{1}{4} 
\int_{\delta}^1 d\alpha
 \frac{\partial}{\partial\alpha} 
\langle \tilde G_0^+\Phi_A(-\alpha^{-1})~\Phi_B(-\alpha) 
~  G^+_0\Phi_C(\alpha)~
 \Phi_D(\alpha^{-1}) $$
$$- G_0^+\Phi_A(-\alpha^{-1})~\Phi_B(-\alpha) 
~  \tilde G^+_0\Phi_C(\alpha)~
 \Phi_D(\alpha^{-1}) 
 ~ \rangle. $$

Comparing with (\ref{As}), one finds
\begin{equation}
{\cal A}_s + {\cal A}_t = 
-\frac{1}{4} 
\int_0^1 d \alpha 
\langle~(\int d^2 z \mu_{\alpha}(z,\bar z) J^{--}(z)) 
\label{Asum}
\end{equation}
$$( G^+_0\Phi_A(-\alpha^{-1})~\tilde G^+_0\Phi_B(-\alpha) + 
\tilde G^+_0\Phi_A(-\alpha^{-1})~ G^+_0\Phi_B(-\alpha) ) $$
$$( G^+_0\Phi_C(\alpha)~\tilde G^+_0\Phi_D(\alpha^{-1}) + 
\tilde G^+_0\Phi_C(\alpha)~ G^+_0\Phi_D(\alpha^{-1}) ) \rangle $$
$$- 
\frac{1}{4} 
\langle \tilde G_0^+\Phi_A(-\delta^{-1})~\Phi_B(-\delta) 
~  G^+_0\Phi_C(\delta)~ \Phi_D(\delta^{-1}) $$
$$-  G_0^+\Phi_A(-\delta^{-1})~\Phi_B(-\delta) 
~  \tilde G^+_0\Phi_C(\delta)~ \Phi_D(\delta^{-1})~\rangle. $$
Note that we have used the ``cancelled propagator'' argument of
\cite{Polchinski} to ignore the surface term coming from $\alpha=1$,
but this argument cannot be used to ignore the surface term coming
from $\alpha=\delta$.

\subsection{Computation of quartic vertex contribution}

It will now be shown that this surface term from $\alpha=\delta$
is precisely cancelled by the contribution ${\cal A}_q$
from the quartic vertex. Using the action of (\ref{vertices}) and
including only contributions with the cyclic order $[A,B,C,D]$,
\begin{equation}
{\cal A}_q= -\sum_{cyclic}\frac{1}{24} \langle~(\Phi_A (-\delta^{-1}) ~
G_0^+ \Phi_B (-\delta)~ - ~G_0^+ \Phi_A (-\delta^{-1})
~\Phi_B (-\delta) )
\label{Ac}
\end{equation}
$$
(\Phi_C(\delta) ~
\tilde G_0^+ \Phi_D (\delta^{-1})~ -~ \tilde G_0^+ \Phi_C(\delta)
~\Phi_D (\delta^{-1}) )~\rangle$$
where $\sum_{cyclic}$ denotes a sum over the four cyclic permutations
of $[A,B,C,D]$. Using properties of the cyclic sum, the sixteen terms in
${\cal A}_q$ can be reordered as
$${\cal A}_q =
 -\sum_{cyclic}\frac{1}{24} \langle~2\Phi_A ~
G_0^+ \Phi_B ~
\Phi_C ~
\tilde G_0^+ \Phi_D  $$
$$  -\Phi_A  ~
G_0^+ \Phi_B ~
 \tilde G_0^+ \Phi_C
~ \Phi_D 
~-~ 
  G_0^+ \Phi_A ~
\Phi_B 
~\Phi_C ~
\tilde G_0^+ \Phi_D ~\rangle $$
where we have suppressed the $z$ location of the four external strings. Note
that the quartic vertex is invariant under cyclic rotations of these
$z$ locations.
Deforming the contour of $\tilde G^+$ off of 
$\Phi_C$ in the second term, 
one gets
\begin{equation}
{\cal A}_q =
 -\sum_{cyclic}\frac{1}{24} \langle~3\Phi_A ~
G_0^+ \Phi_B ~
\Phi_C ~
\tilde G_0^+ \Phi_D 
\label{used}
\end{equation}
$$  -\tilde G_0^+\Phi_A  ~
G_0^+ \Phi_B ~
 \Phi_C
~ \Phi_D 
~-~ 
  G_0^+ \Phi_A 
~\Phi_B 
~\Phi_C ~
\tilde G_0^+ \Phi_D ~\rangle .$$
But the last two terms of (\ref{used}) cancel each other after
summing over cyclic permutations, so
\begin{equation}
{\cal A}_q =
 -\sum_{cyclic}\frac{1}{8} \langle~\Phi_A  ~
G_0^+ \Phi_B ~
\Phi_C ~
\tilde G_0^+ \Phi_D ~\rangle
\label{feq}
\end{equation}
$$= -\frac{1}{8} \langle
~
 \Phi_A  ~
G_0^+ \Phi_B ~
\Phi_C ~
\tilde G_0^+ \Phi_D ~-~
 \Phi_A  ~
\tilde G_0^+ \Phi_B ~
\Phi_C ~
G_0^+ \Phi_D~\rangle ~+\frac{X}{8} $$
where 
$$X= 
\langle \tilde G_0^+\Phi_A  ~
\Phi_B ~
G^+_0\Phi_C ~
\Phi_D ~-~
G_0^+\Phi_A  ~
\Phi_B ~
\tilde G_0^+\Phi_C ~
\Phi_D ~\rangle.$$

Deforming the $G^+$ contour off of $\Phi_B$ in the first term and
then deforming the $\tilde G^+$ contour off of $\Phi_D$, 
one finds
$${\cal A}_q =
 -\frac{1}{8} \langle
~
- G_0^+ \Phi_A  ~
\Phi_B ~
\Phi_C ~
\tilde G_0^+ \Phi_D $$
$$~-~
\Phi_A~\Phi_B ~G_0^+\Phi_C ~\tilde G_0^+\Phi_D 
~-~ \Phi_A  ~
\tilde G_0^+ \Phi_B ~
\Phi_C ~
G_0^+ \Phi_D~\rangle ~+\frac{X}{8} $$
$$= 
 -\frac{1}{8} \langle
~ G_0^+ \Phi_A  ~
\tilde G_0^+ \Phi_B ~
\Phi_C ~
\Phi_D ~-~
\Phi_A~\tilde G_0^+\Phi_B ~G_0^+\Phi_C ~\Phi_D  $$
$$
~-~ \Phi_A  ~
\tilde G_0^+ \Phi_B ~
\Phi_C ~
G_0^+ \Phi_D~\rangle ~+\frac{X}{4}. $$
Finally, deforming the $G_0^+$ contour off of $\Phi_D$ in the last term,
one finds 
\begin{equation}
{\cal A}_q = \frac{X}{4} .
\label{Acresult}
\end{equation}

So combining (\ref{Acresult}) with (\ref{Asum}), 
\begin{equation}
{\cal A} = {\cal A}_s + {\cal A}_t + {\cal A}_q
=
-\frac{1}{4} 
\int_0^1 d \alpha 
\langle~(\int d^2 z \mu_{\alpha}(z,\bar z) J^{--}(z)) 
\label{result}
\end{equation}
$$( G^+_0\Phi_A(-\alpha^{-1})~\tilde G^+_0\Phi_B(-\alpha) + 
\tilde G^+_0\Phi_A(-\alpha^{-1})~ G^+_0\Phi_B(-\alpha) ) $$
$$( G^+_0\Phi_C(\alpha)~\tilde G^+_0\Phi_D(\alpha^{-1}) + 
\tilde G^+_0\Phi_C(\alpha)~ G^+_0\Phi_D(\alpha^{-1}) ) \rangle $$
$$=
\int_0^1 d \alpha 
\langle~(\int d^2 z \mu_{\alpha}(z,\bar z) G^{-}(z)) 
$$
$$ \Phi_A(-\alpha^{-1})~ G^+_0\Phi_B(-\alpha) ~ 
 G^+_0\Phi_C(\alpha)~\tilde G^+_0\Phi_D(\alpha^{-1})  ~
\rangle $$
where we have used contour deformations of $G^+$ and $\tilde G^+$ 
to write all four terms
in the same form.
Note that there are no surface term contributions coming from these
deformations due to the ``cancelled propagator'' argument \cite{Polchinski}
which can be used when $\alpha=0$ or $\alpha=1$.

So we have proven our claim that, after including the contribution from
the quartic vertex of the action of (\ref{action}), the  second-quantized
prescription agrees with the first-quantized prescription
for on-shell tree-level four-point open superstring
scattering amplitudes. 

{\bf Acknowledgements:} The authors would like to thank CNPq grant
300256/94-9 and FAPESP grant 96/12280-0 for financial support.


\begin{thebibliography}{99}
\bibitem{wendt} C. Wendt, Nucl. Phys. {\bf B314} (1989) 209.
\bibitem{witten2} E. Witten, Nucl. Phys. {\bf B276} (1986) 291.
\bibitem{preit} C.R. Preitschopf, C.B. Thorn, and S. Yost,
Nucl. Phys. {\bf B337} (1990) 363;
I. Ya. Aref'eva, P.B. Medvedev, and A.P. Zubarev,
Nucl. Phys. {\bf B341} (1990) 464;
I. Ya. Aref'eva and P.B. Medvedev,
Phys. Lett. {\bf B202} (1988) 510;
N. Berkovits, M. Hatsuda and
W. Siegel, Nucl. Phys. {\bf B371} (1992) 434, hep-th/9108021.
\bibitem{GF} J. Greensite and F.R. Klinkhamer, Nucl. Phys. {\bf B291}
(1987) 557; J. Greensite and F. Klinkhamer, Nucl. Phys. {\bf B304} (1988) 108.
\bibitem{field} N. Berkovits, Nucl. Phys. {\bf B450} (1995) 90. 
\bibitem{witten1} E. Witten, Nucl. Phys. {\bf B268} (1986) 253.
\bibitem{topo} N. Berkovits and
C. Vafa, Nucl. Phys. {\bf B433} (1995) 123, hep-th 9407190.
\bibitem{Donaldson} S. Donaldson, Proc. Lond. Math. Soc. {\bf 50} (1985) 1;
V.P. Nair and J. Schiff, Phys. Lett. {\bf B246} (1990) 423.
\bibitem{four} N. Berkovits, Nucl. Phys. {\bf B431} (1994) 258,
hep-th 9404162.
\bibitem{leite} N. Berkovits and M.M. Leite, Phys. Lett. {\bf B454} (1999) 38,
hep-th 9812153.
\bibitem{giddings} S. Giddings, Nucl. Phys. {\bf B278}  (1986) 242.
\bibitem{FMS} D. Friedan, E. Martinec and S. Shenker, Nucl. Phys. {\bf B271}
(1986) 93.
\bibitem{Polchinski} J. Polchinski, {\bf String Theory, vol.1},
Cambridge University Press (1998).


 
\end{thebibliography}
\end{document}